\documentclass[preprint,showpacs,showkeys,amsmath,amssymb]{revtex4}
\usepackage{graphicx,caption2,subfigure}
\begin{document}
\title{The $p_T$-Spectra of Some Non-pion Secondaries in High Energy
$NN$ to $NA/AA$ Collisions and the Combinational Approach}
\author{Bhaskar De}
\email{bhaskar_r@isical.ac.in}
\author{S. Bhattacharyya}
\email{bsubrata@isical.ac.in (Communicating Author).}
\affiliation{Physics and Applied Mathematics Unit(PAMU),\\
Indian Statistical Institute, Kolkata - 700108, India.}
\date{\today}

\begin{abstract}
In continuation of our perusal of the studies on transverse
momentum spectra for the main varieties of secondaries from a
consistent and comprehensive phenomenological approach, we propose
to take up here --- after a successful completion of reporting in
detail the results (Ref.\cite{De2} in the text) on our analyses of
the $p_T$-spectra of pions --- the studies specially on production
of kaons, protons and antiprotons in several proton-induced and
nucleus-involved collisions at high energies. The measured data on
inclusive cross sections of kaons, protons and antiprotons from
the various published sources have here been assorted first. Next,
data on the $p_T$-spectra of the specific secondaries produced in
$PP$ and $P\bar{P}$ reactions have been scanned and analyzed
with the help of Hagedorn's model(HM). Thereafter a connector,
named here the combinational approach(CA), has been constructed
and used to analyze the data on $p_T$-spectra of some major
category of non-pion secondaries produced in nucleus-nucleus($AA/AB$)
collisions at high energies. And these exercises have, finally,
led to the modestly successful interpretations of a wide band of
data with the revelation of some insightful physical aspects of
high energy interactions. The limitations of the approach have
also been precisely pointed out in the end.
\end{abstract}
\keywords{Relativistic Heavy Ion Collision, Inclusive Cross Section.}
\pacs{25.75.-q, 13.60.Hb.}
\maketitle

\newpage

\section{Introduction}

The nature of transverse momentum\cite{Hwa2,Zhang1} spectra is quite interesting for
the following reasons: Firstly, in the form of invariant cross
section in terms of $p_T$, it offers a ready, reliable and very
basic observable for both measurements, and for theorization to be
attempted. Secondly, the $p_T$-spectra throw light on the particle
production mechanism on the whole; they have also impact and
indirect reflections on the predicted features of what is normally
projected as the signatures of the conjectured quark-gluon
plasma(QGP). Thirdly and finally, the expressions for invariant $p_T$-spectra lead us to define some other physically significant observables in the domain of particle physics, like the average multiplicity, average transverse momentum etc.
\par
After pions, kaons of all varieties (with $K^\pm$ and
$K^0/\bar{K}^0$) constitute the second most abundant species among
the particle secondaries which are produced in multiparticle
processes of hadron-hadron, hadron-nucleus or nucleus-nucleus
interactions at high and very high energies. Secondly, kaons are
the lightest `strange' particles for which they have a special
status. Besides, kaons are supposed to bear, as told above, some
strong reflections on what are viewed as the QGP diagnostics.
Lastly, the kaon-pion ratios and their energy-dependences play a
very significant role in both the particle physics and non-accelerator physics sectors. So, analyzing the kaon production characteristics, especially the $p_T$-spectra, in the light of any model assumes a high degree of importance and comes as a challenging task as well. Furthermore,
the production of proton-antiprotons, the first family-set of
baryonic category, is also important from various physical
considerations. The nature of the ratios $\bar{P}/P$ has a strong
bearing on astroparticle physics. Besides, the generalized
antiparticle to particle ratios are of importance in the study of
matter-antimatter symmetry in the universe, since they provide the
relative abundance of antiparticle productions\cite{Staszel1}.
\par
Very recently, spurts of data on the various aspects of particle
production from the CERN-SPS and the RHIC-BNL, studying high energy interaction properties in the collisions between $PbPb$ and $AuAu$ respectively, are available. And providing interpretations for both the previous measurements and these oncoming blizzards of new data is really a very significant work.
\par
Our objective here is to try to describe and/or explain the nature
of the transverse momentum ($p_T$)-spectra of kaons, protons and
antiprotons in a large variety of proton-induced and
nucleus-induced high energy interactions with the help of a
combinational approach(CA), called also the grand combination of models (GCM). The present work is prompted by our previous successes in
analyzing the $p_T$-spectra and rapidity-spectra\cite{De2,De1} of
pions alone in a host of nuclear collisions at high and superhigh
energies. In fact, taking up the task of interpreting the
$p_T$-spectra of a set of non-pion secondaries produced in nuclear
collisions becomes the logical imperative and the natural
follow-up in maintaining a sequence of our studies and with the
model of our choice in a self-complete way. In real terms, while proceeding with the present work, we would follow the trajectory of the Ref.\cite{De2} here more closely than the other one, as in the former the approach to the studies on $p_T$-spectra has been rationalized in a better way and much more consistent manner. We have consciously and carefully avoided here to adopt operationally any of the known popular/standard models and have also tried to escape, as far as possible, the buzzwords related to them. And we have singularly been guided by the results of experimental measurements on the nature of $p_T$-spectra alone in a very dispassionate and objective manner. 
\par
This paper is organized here as follows. In the next section (section
II) we give the outline of the combinational approach which is to
be taken up for this study and the sketch of the physical
perspective which prompted us to proceed in the stated direction.
In section III we present the essential steps of the methodology
of our work, the results of model-based calculations and some
brief discussion on the results obtained by this combinational
approach. The last section is reserved for summing up the
conclusions and pointing out the shortcomings of the approach adopted here.

\section{The Basic Approach and the Physics Perspective}

Following the suggestion of Faessler\cite{Faessler1} and the work
of Peitzmann\cite{Peitzmann1} and also of Schmidt and
Schukraft\cite{Schmidt1}, we propose here a generalized empirical
relationship between the inclusive cross-section for production of $K^\pm$, $P$ and $\bar{P}$ in nucleon(N)-nucleon(N) collision and that for
nucleus(A)-nucleus(A) collision as given below:

\begin{equation}
\displaystyle{E\frac{d^3  \sigma}{dp^3} ~ (AB \rightarrow Q X) ~
\sim ~ (A.B)^{\phi(y, ~ p_T)} ~ E\frac{d^3  \sigma}{dp^3} ~ (PP
\rightarrow Q X) ~,} \label{e1}
\end{equation}

where Q stands for $K^\pm$, $P$, $\bar{P}$ and $\phi(y, ~ p_T)$
could be expressed in the factorization form, $\phi(y, ~ p_T) =
f(y) ~ g(p_t)$.
\par
While investigating a specific nature of dependence of
the two variables($y$ and $p_T$), either of them is assumed to
remain averaged or with definite values. Speaking in clearer
terms, if and when $p_t$-dependence is studied by experimental
group, the rapidity factor is integrated over certain limits
and is absorbed in the normalization factor.So, the formula turns into

\begin{equation}
\displaystyle{E\frac{d^3 \sigma}{dp^3} ~ (AB \rightarrow Q X) ~
\sim ~ (A.B)^{g(p_T)} ~ E\frac{d^3  \sigma}{dp^3} ~ (PP
\rightarrow Q X) ~,} \label{e2}
\end{equation}

The main bulk of work, thus, converges to the making of an appropriate
choice of form for $g(p_T)$. And the necessary choices are to be made
on the basis of certain premises and physical considerations which do
not violate the canons of high energy particle interactions.
\par
Let us now tentatively propose that the expressions for inclusive
cross-section of non-pion secondaries in proton-proton scattering
at high energies in eqn.(2) could also be chosen in the same form
as that was suggested for pions by Hagedorn\cite{Hagedorn1}:
\begin{equation}
\displaystyle{ E\frac{d^3  \sigma}{dp^3} ~ (PP \rightarrow Q X)
~ = ~ C_1 ~ ( ~ 1 ~ + ~ \frac{p_T}{p_0})^{-n} ~ ,} \label{e3}
\end{equation}

where $C_1$ is the normalization constant, and $p_o$, $n$ are
interaction-dependent chosen phenomenological parameters for which
the values are to be obtained by the method of fitting. Their
$\sqrt{s}$-dependences are here proposed to be given by the
following formulations:

\begin{equation}
\displaystyle{ p_0(\sqrt{s}) ~ = ~ a ~ + ~ \frac{b}{\sqrt{s} ~
\ln({\sqrt{s}})}}
\label{e4}
\end{equation}
and
\begin{equation}
\displaystyle{ n(\sqrt{s}) ~ =  ~ \acute{a} ~ + ~
\frac{\acute{b}}{\ln^2({\sqrt{s}})}}
\label{e5}
\end{equation}

 where $a$, $b$, $\acute{a}$ and $\acute{b}$ are four constants.
The $\sqrt{s}$-dependence of $p_0$ and $n$ would be shown later
diagrammatically in the text and their data-base would also be
indicated. The nature and significance of these parameters could
be appreciated from the work of Hagedorn\cite{Hagedorn1}, and
those of Bielich et al\cite{Bielich1} and Albrecht et
al\cite{Albrecht1}.
\par
The final working formula for the nucleus-nucleus collisions is
now being proposed here in the form given below:

\begin{equation}
\displaystyle{E\frac{d^3  \sigma}{dp^3} ~ (AB \rightarrow Q X) ~
\approx ~ C_2 ~ (A.B)^{(\epsilon ~ + ~ \alpha  p_T  ~ - ~ \beta
p_T^2)} ~ E\frac{d^3 \sigma}{dp^3} ~ (PP \rightarrow Q X) ~,}
\label{e6}
\end{equation}

with $ g(p_T) ~ = ~ (\epsilon ~ + ~ \alpha  p_T  ~ - ~ \beta
p_T^2)$, wherein this quadratic form of parametrization is
suggested here tentatively by us for analyzing data and testing
its efficacy, if any and is hereafter called De-Bhattacharyya
parametrization(DBP). In the above expression $C_2$ is the
normalization term which has a dependence either on the rapidity
or on the rapidity-density of the produced secondary; $\epsilon$, $\alpha$ and $\beta$ are constants for a specific set of projectile and target.
\par
Earlier experimental works\cite{Albrecht1,Antreasyan1,Aggarwal1}
showed that $g(p_T)$ is less than unity in the $p_T$-domain,
$p_T<1.5$ GeV/c. Besides, it was also observed that the parameter
$\epsilon$, which gives the value of $g(p_T)$ at $p_T=0$, is also
less than one and this value differs from collision to collision.
The other two parameters $\alpha$ and $\beta$ essentially
determine the nature of curvature of $g(p_T)$. However, in the
present context, precise determination of $\epsilon$ is not
possible for the following understated reasons:
\par
(i) To make our point let us recast the expression for (6) in the
form given below:
\begin{equation}
\displaystyle{E\frac{d^3\sigma}{dp^3}(AB \rightarrow Q X) ~
\approx ~ C_2 ~ (A.B)^\epsilon ~ (A.B)^{(\alpha p_T - \beta
p_T^2)} ~ ( ~ 1 ~ + \frac{p_T}{p_0} ~ )^{-n}}\label{e7}
\end{equation}
Quite obviously, we have adopted here the method of fitting. Now,
in eqn.(7) one finds that there are two constant terms $C_2$ and
$\epsilon$ which are neither the coefficients nor the exponent
terms of any function of the variable, $p_T$. And as $\epsilon$ is
a constant for a specific collision at a specific energy, the
product of the two terms $C_2$ and $(A.B)^\epsilon$ appears as
just a new constant. And, it will not be possible to obtain
fit-values simultaneously for two constants of the above types
through the method of fitting.
\par
(ii) From eqn.(2) the nature of $g(p_T)$ can easily be determined
by calculating the ratio of the logarithm of the ratios of
nuclear-to-$PP$ collision and the logarithm of the product $AB$.
Thus, one can measure $\epsilon$ from the intercept of $g(p_T)$
along y-axis as soon as one gets the values of
$E\frac{d^3\sigma}{dp^3}$ for both $AB$ collision and $PP$
collision at the same c.m. energy. In the present study we have
tried to consider the various collision systems in as many number
as possible. To do so, we have to consider the data on normalized
versions of $E\frac{d^3\sigma}{dp^3}$ for many collision systems
for which clear $E\frac{d^3\sigma}{dp^3}$-data were not available
to us. Furthermore, from these normalized versions we can/could
not extract the appropriate values of $E\frac{d^3\sigma}{dp^3}$ as
the normalization terms, total inclusive
cross-sections$(\sigma_{in})$ etc., for these collision systems
cannot always be readily obtained. Besides, it will also not be
possible to get readily  the data on inclusive spectra for $PP$
collisions at all c.m.energies, like e.g., at $\sqrt{s}=17.8
GeV$(c.m. energy of $Pb+Pb$ collision).
\par
In order to sidetrack these difficulties and also to build-up an
escape-route, we have concentrated almost wholly here to the values of
$\alpha$ and $\beta$ for various collision systems; and the effects
of $C_2$ and $\epsilon$ have been compressed to a single constant
term $C_3$. Hence, the final expression becomes,
\begin{equation}
\displaystyle{E\frac{d^3\sigma}{dp^3}(AB \rightarrow Q X) ~
\approx ~ C_3 ~ (A.B)^{(\alpha p_T - \beta p_T^2)} ~ ( ~ 1 ~ +
\frac{p_T}{p_0} ~ )^{-n}}\label{e8}
\end{equation}
with $C_3 = C_2 (A.B)^\epsilon$.
\par
The exponent term $\alpha p_T - \beta p_T^2$ obviously represents
here $[g(p_T)-\epsilon]$ instead of $g(p_T)$ alone. Thus, after
obtaining fit-values of $\alpha$ and $\beta$, if
$[g(p_T)-\epsilon]$ are plotted for various collision systems, all
the curves would originate from a single point, i.e. origin; and
the systems and processes are then really comparable. In other
words, in this convenient way we could study and check the scaling
characteristics of $g(p_T)$ with respect to the collision systems.
\par
The expression(8) given above is the physical embodiment of what
we have termed to be the grand combination of models(GCM) or the
combinational approach(CA) that has been applied here. Firstly,
the results of $PP$ scattering are obtained in the above on the
basis of an assumption of validity of eqn.(3) for even the
non-pion secondaries, though originally the Hagedorn's model(HM)
was proposed only for pion secondaries. Secondly, the route for
converting the results for the non-pion secondaries produced in
$NN$ to $NA$ or $AB$ collisions is built up by the Peitzmann's
approach(PA) represented by expression(2). Thirdly, the further
input is the De-Bhattacharyya parametrization for the nature of
the exponent. Thus, the GCM or the CA provides the combination of
HM, PA and the DBP, and thus constitutes a single whole. And this
whole is both useful and economical; it is economical in the sense
that it can accommodate vast wealth of data with only three
arbitrary parameters. Compared to the theoretical apparatus
provided by the standard model(SM)\cite{Ellis1}, this number is
quite low.
\par
And the choice of this quadratic form is not altogether a
coincidence. In dealing with the EMC effect in the lepton-nucleus
collisions, one of the authors here(SB)\cite{SB1}, made use of a
polynomial form of $A$-dependence with the variable $x_F$(Feynman
Scaling variable). This gives us a clue to make a similar choice
with both $p_T$ and $y(\eta)$ variable(s) in each case separately.
In the recent times, De-Bhattacharyya parametrization is being
extensively applied to interpret the measured data on the various
aspects\cite{De2} of the particle-nucleus and nucleus-nucleus
interactions at high energies. In the recent past Hwa et.
al.\cite{Hwa1} also made use of this sort of relationship in a
somewhat different context. The underlying physics implications of
this parametrization stem mainly from the expression(6) which
could be identified as a clear mechanism for switch-over of the
results obtained for nucleon-nucleon($PP$) collision to those for
nucleus-nucleus interactions at high energies in a direct and
straightforward manner. The polynomial exponent of the product
term on $AB$ takes care of the totality of the nuclear effects.
\par
The individual model(s), seen in a split manner, is(are) certainly
not new; but the combination with the proposed two-factor
quadratic $p_T$-dependence of the exponent (called DBP in the
text) for $AB$ or $AA$ interactions at high energies is testably a
new proposition offered by us. Besides, the probing of the role of
the Hagedorn's model in understanding the production of non-pion
secondaries in high energy collisions is another virgin feature in
our work. The testing of this mechanism suggested by us is with
the views of (i) presenting an integrated approach toward
production of various kind of particle-secondaries in high energy
collisions and (ii) providing a unified outlook to particle
production in all varieties of particle-particle, particle-nucleus
or nucleus-nucleus interactions at high energies. Quite
understandably, only by intensive analysis of data on several/all
sets of collisions somewhat successfully such claims could, at
all, be made, and/or later be justified. The trends that emerge
from this extensive data-analysis with this grand combination of
models(GCM) has been given in the next section.
\par
Obviously,in its approach and method, this work is undoubtedly a
continuation of one of our previous works\cite{De2}. But, even at
the risk of being a bit repetitive, we have to proceed and carry on.
Because, without these special efforts at studying on a
case-to-case basis for each of the specific variety of the
secondaries, we cannot reasonably arrive at any definitive
conclusion and also cannot confirm the physical basis, if any, of
our work. And the two points made very precisely in the preceding
paragraph represent the strong and crucial aspects of our physics
motivation behind this proposed formalism. A point is to be made.
Our intention here, is surely not just piling up of large bulk of
data in a purposeless way; rather we would like to make the
fullest possible utilization of them for checking the proposed GCM
as extensively as possible with the widest range of available data
on diverse set of reactions. Because, such rigorous and intensive
checkings only could lead to meaningful and valid conclusions, if
any. And, in our opinion, this has helped us to substantiate, to a
considerable degree, the functional efficacy and strength of the
proposed phenomenological model which is called here the
combinational approach. In fact, the revelation of the potency and
strength of this phenomenological approach is the other most striking feature which we would surely like to emphasize here.
\par
It is obvious that there are two phenomenological parameters in
$g(p_T)$ in expression(8) which need to be physically explained
and/or identified. In compliance with this condition we offer the
following physical explanations for the occurrence of all these
factors. The particle-nucleus or nucleus-nucleus collisions at
high energies subsequently gives rise to an expanding blob or
fireball with rising temperature. In real and concrete terms this
stage indicates the growing participation of the already-expanded
nuclear blob. As temperature increases at this stage, the emission
of highly energetic secondaries(which are mostly peripheral
nucleons or baryons) with increasing transverse momentum is
perfectly possible. The coefficient $\alpha$ addresses this
particularity of the natural event; and this is manifested in the
enhancement of the nuclear contribution with the rise of the
transverse momentum. Thereafter, there is a turnabout in the state
of reality. After the initial fractions of seconds, the
earlier-excited nuclear matter starts to cool down and there is a
clear natural contraction at this stage as the system suffers a
gradual fall in temperature. Finally, this leads to what one might
call `freeze-out' stage, which results in extensive hadronization,
especially in production of hadrons with very low transverse
momentum. In other words, the production of large-$p_T$ particles
at this stage is lowered to a considerable extent. This fact is
represented by the damping or attenuation term for the production
of high-$p_T$ particles. The factor $\beta$ with negative values
takes care of this state of physical reality. Thus the function
denoted by $g(p_T)$ symbolizes the totality of the features of the
expansion-contraction dynamical scenario in the after-collision
stage. This interpretation is, at present, only of very suggestive nature.
However, let us make some further clarifications.
\par
The physical foundation that has here been attempted to be built
up is inspired by thermodynamic pictures, whereas the quantitative
calculations are based on a sort of pQCD-motivated power-law
formula represented by eqn.(6). This seems to be somewhat
paradoxical, because it would be hard to justify the hypothesis of
local thermal equilibrium in multihadron systems produced by high
energy collisions in terms of successive collision of the
QCD-partons(like quarks and gluons) excited or created in the
course of the overall process. Except exclusively for central
heavy ion collisions, a typical parton can only undergo very few
interactions before the final-state hadrons `freeze out', i.e.
escape as free particles or resonances. The fact is the hadronic
system, before the freeze-out starts, expands a great deal -- both
longitudinally and transversally -- while these very few
interactions take place\cite{Hove1}. But the number of parton
interactions is just one of the several other relevant factors for
the formation of local equilibrium. Of equal importance is the
parton distribution produced early in the collision process. This
early distribution is supposed to be a superposition of collective
flow and highly randomized internal motions in each space cell
which helps the system to achieve a situation close to the
equilibrium leading to the appropriate values of collective
variables including concerned and/or almost concerned quantities.
The parameter $\alpha$ in expression(6) is somehow related to the  measure of the ratio of the net binary collision number to the total permissible number among the constituent partons in the pre-freeze out expanding stage identified to be a sort of explosive
`detonation'\cite{Hove1} stage. This is approximated by a
superposition of collective flow and thermalized internal motion,
which is a function of hadronic temperature manifested in the
behaviour of the average transverse momentum. The post freeze-out
hadron production scenario is taken care of by the soft
interaction which is proportional\cite{Aggarwal1,Li1} to the
number of participant nucleons, $N_{part}$, according to almost
any variety of wounded nuclear model. The factor $\beta$, we
conjecture, might have a relationship with the ratio of the actual participating nucleons to the total number of maximum allowable(participating) nucleons. In fact, this sort of physical explanations is reproduced here from some of our previous works\cite{De3}.

\section{Procedural Steps, Results and Discussion}

At the very start we study the $p_T$-spectra for $K^{\pm}$, $P$
and $\bar{P}$ inclusive production in $PP$ collisions at several
energies and try to fit the expression in the formula(3) given in
the previous section. The graphs are shown in the several diagrams
in Fig.1 for kaons, protons and antiprotons. The fits for the
average kaon yield and antiproton production in $P\bar{P}$
reactions have also been obtained(Fig.2) on the basis of
Hagedorn's model(eqn.(3)). The same could not be done for proton
production in $P\bar{P}$ reactions due to unavailability of the
measured experimental data in this particular collisions. The
obtained fitted-values of the parameters, $p_0$ and $n$, for
different secondaries produced in $PP/P\bar{P}$ collisions at
different energies have been displayed in Table-I - Table-V. The
graphs for $\sqrt{s}$-dependences of $p_0$ and $n$, as proposed in
eqn.(4) and eqn.(5), have been depicted in Fig.3 and the necessary
values of the parameters, $a$, $\acute{a}$, $b$ and $\acute{b}$
have been presented in Table-VI. While plotting graphs we have
obviously assumed that at very high energies the proton-proton and
proton-antiproton collisions could be treated at par and with the
acceptance of equivalence between each other.
\par
Hereafter, in studying the nature of $p_T$-spectra in all nucleon-nucleus and nucleus-nucleus collisions, the interaction energy in all cases of
nucleus-induced reactions is invariably converted first into the
c.m. system values, that is expressed in $\sqrt{s_{\rm NN}}$, then
the values of $p_0$ and $n$ are picked up from the graphical plots drawn already and shown by Fig.3. While analyzing nucleus-dependence
with expression(8) and trying with the fit parameters $C_3$,
$\alpha$ and $\beta$, we have inserted these extracted values of
$p_0$ and $n$ for $PP$ cross section term occurring in the
expression(8) given above.
\par
However, the plots presented in Fig.3 deserve special mention from
phenomenological points of view. The $p_0$ and $n$ values for
production of the same secondary at various energies
($\sqrt{s}$-values) show very slow fall-off with energy in these
graphs. The nearness of values of $p_0/n$ signals a march towards
a steady state of hadronization. Once the natures are established
by the theoretical procedures adopted for either $PP/P\bar{P}$
reactions at high or very high energies, we would be compelled,
for the sake of consistency, to reduce/minimize the degree of
arbitrariness in choosing values of $p_0$ and $n$ for
nucleon-nucleus or nucleus-nucleus collisions. As soon as the
values of $\sqrt{s_{NN}}$ is given for a specific interaction, the
values of $p_0$ and $n$ are to be obtained from the theoretical
plots of them shown in Fig.3(a), Fig.3(b) and Fig.3(c) for
different secondaries and/or for different collision(s). The
diagrams given in Fig.4 to Fig.8 do actually provide the nature of
collected data and also the theoretical descriptions for diverse
reactions indicated by the labels in each diagram. The solid
curves represent the results arrived at on the basis of the new
combination of models(NCM) and the use of De-Bhattacharyya
parametrization(DBP). And the totality of this combination of NCM
and DBP is called Grand Combination of Models(GCM). And the tables
(displayed by Table-VII to Table-IX) provide the necessary
parameter values that have entered into each calculation of the
theoretical solid curves in all the diagrams of Fig.4-Fig.8 for
various reactions. In some cases, due to lack of availability of
clear proton data, we have used the net proton($P-\bar{P}$) data
with the assumption of the near equivalence and as an approximate
measure. The diagrams presented in Fig.9(a) to Fig.9(c) are full
of physical import and convey the message of mass number scaling
of nuclear reactions.
\par
The deviations, whenever and wherever if any, from this
observation could be attributed, in the main, to the following few
points: (i) limits of uncertainties in the measurements; (ii) some
unavoidable approximations in theoretical calculations; and (iii)
availability, in some cases, of only too sparse and rare data on which
reliability is certainly questionable.
\par
Obviously, we present here the GCM-based analyses of extensive
sets of data on some of the observables measured by the various
experimental groups in high energy particle and nuclear physics.
The study has, thus, given rise to some crucial observations with
revelations of following systematic trends: (a) For the non-pion
secondaries produced in nucleus-nucleus collisions at high
energies the nature of fits is not as good as in $PP$ reaction,
especially at low momentum region, and especially for
$E\frac{d^3N}{dp^3}$ vs. $p_T$ plots. The reason(s) for this
discrepancy would be discussed later in the last section. (b) The
fit parameters $p_0$ and $n$ for production of secondary kaon(s)
and protons are, by ascription, different in the basic $PP$
reactions. And this is consistent with the theoretical expectations. The factor $p_0/n$ has, we recognize, a relationship with the slope parameter of the $p_T$-spectra. And the slope parameters for the lighter and heavier particles, one knows, are not the same. But, quite interestingly, the nature of $\sqrt{s}$-dependences of $p_0$ and $n$ values of all three
species, kaons, protons and antiprotons, as shown in
Fig.3(a)-3(b), are qualitatively quite alike.(c) It is quite
evident from the Tables VII, VIII, and IX that the values of
$\alpha$ and $\beta$ are too close. This, in our opinion, reflects
the basic fact that the nuclear effects are quite finite and
certainly not too pronounced even with the heaviest of the nuclei
and the highest available energies.

\section{Final Comments and Conclusions}

In the end let us summarize our main findings from the present
study in the following way:
\begin{enumerate}
\item[(i)] Quite naturally and predictably, the $\chi^2/ndf$
values shown in Table-VII-IX present, at times, sharply contrasting nature, because of either
very sparse available data and that too for a narrow region of $p_T$-values, or large ranges
of errors and uncertainties in the measurements. \item[(ii)]
Still, on the whole, the approach describes modestly well the large bulk of data
on kaon-antikaon and proton-antiproton production phenomena in
high energy nucleon-nucleus and nucleus-nucleus collisions.
\item[(iii)]Related with production of $K^{\pm}$ and $P/\bar{P}$,
the c.m. energy-dependences of the parameter values occurring in
the basic $PP$ reactions are of the same hyperbolic nature as in
the cases of pions. This element of observation on similarity
aspect of $K^\pm$ and $P/\bar{P}$ with pions is certainly a new
finding here made by us, as Hagedorn's early work\cite{Hagedorn1} was made
exclusively on pions.  \item[(iv)] Besides, the entire
energy-dependences, even in purely nuclear collisions, are
manifested by only the basic $PP$ interaction; the nuclear
geometry exhibits no separate energy behaviour. This is quite
significant in the sense that it might be a great hint to and
confirmation of the enormous importance of $PP$ interaction even
in understanding the nature of heavy ion collisions. \item[(v)]
The observation of the close values of $\alpha$ in various sets of
nuclear collisions with varying pair of projectile and target and
at different energies is another interesting finding. It indicates that
enhancement due to nuclear contribution could never assume too
large values, even with the heaviest of the nuclei, though the
parameters are uniformly non-zero for each and every collision and
the values are limited. This does, in effect, imply the very
finite degree of the nuclear effects. The constraint would appear
more severe and acute, if the diminutive role of the $\beta$
factor is reckoned with. This might be attributed to the large
baryon stopping effect in heavy ion collisions. 
\par
Besides, the small positive $\alpha$-values in any collission represent physically the nuclear enhancement --- though certainly not of `anomalous' nature --- called Cronin effect. secondly, the negative $\beta$-values might explain the suppressions of productions of hadrons at relatively large values of the transverse momenta.   

\item[(vi)] Even
with some heaviest nuclei, the observations from Fig.9(a) to
Fig.9(c) on the validity of a sort of mass-number($A$) scaling for
production of both strange and non-strange variety of hadrons
constitute an interesting and probeworthy point. Physically, this
might signal the limit to the finite degree of impact parameter
dependence of nuclear collisions. The near-saturation of the
parameter value at and after certain magnitude of $AB$ indicates
the stringent limit on both the number of binary collisions and
the number of participating nucleons in the colliding nuclei.
\item[(vii)] The reason for observation of departure of the fits
from the data on $p_T$-spectra produced in nuclear collisions,
especially at low $p_T$ region, could be ascribed, in the main, to
the following reasons. Firstly, because for the nucleus-involved
collisions, the measurements in the experiments on purely soft
collisions are very difficult, as hard collision effects are
almost unavoidably present there. Secondly, the measurements of
the factor $E\frac{d^3N}{dp^3}$ through the array of several detectors have always an intrinsic simulation-based standard-model
(mainly the HIJING version) component, for which the obtained data are not
the pure products of experimental measurements alone. But the present model can, in no way, accommodate such sorts of superposition effects. \item[(viii)] The
property of `universality' of hadronic, hadronuclear and nuclear
reactions at very high energies is also vindicated by the approach
adopted here.
\end{enumerate}
\par
The implications of all these are obvious. With the series of such
modestly successful ventures, the combinational approach attains a
plateau and assumes a potential to claim the status of an
alternative approach to understand functionally some aspects of
the heavy ion collisions in general, and the nature of
$p_T$/rapidity-spectra in $NA$ or $AB$ collisions in particular.
\par
However, we must not gloss over the gross limitations of the
present approach which are as follows: (i) It is a fact that we could, so
far, offer no measure to estimate quantitatively the values of
$\alpha$ and $\beta$. The hints are, uptil now, of only very
qualitative nature. (ii) Quite admittedly, the physical
explanations outlined here for $\alpha$ and $\beta$ are also of
tentative nature; they have, still, not been identified concretely
with any of the known physical observables in the domain of
collision dynamics for multiple production of hadrons and of
nuclear geometry as well. (iii) The calculations for
particle-nucleus or nucleus-nucleus interaction cases are
absolutely dependent on the apriori and data-based knowledge of
the production characteristics of the same particle species in $PP$
reactions at various energies. Without it, the method is
ineffective and that is a major handicap. (iv) The approach fails
to respond to the very basic query on actual mechanisms for
particle production processes in any collision. (v) And just
because of it, the approach is insensitive to the charge-state of
the specific secondary produced in any high energy interaction and
that is certainly a great difficulty. So, unless these problems
could be remedied to a great extent, we accept that we have no
reasons for complacency.

\newpage

\begin{table}
\begin{ruledtabular}
\caption{Fit Values of $p_0$ and $n$ for $PP \rightarrow (K^++K^-)/2 + X$ at different energies}
\begin{tabular}{ccccccc}
\hline $\sqrt{s}_{NN}(GeV)$ & Relevant & $C_1$ & $p_{0}$(GeV/c) & $n$ & $p_0/n$(GeV/c) & $\chi^2/ndf$ \\
$[$with reference$]$ & collision-specifics & & & & & \\
\hline $23$\cite{Alper1} & &  $12 \pm 3$ & $2.7 \pm 0.9$ & $17 \pm 4$ & 0.16 & 1.427 \\
 $31$\cite{Alper1} & $y_{cm}=0$, & $15 \pm 2$ & $4 \pm 1$ & $22 \pm 4$ & 0.18 & 1.031\\
 $45$\cite{Alper1} & min. bias & $13 \pm 3$ & $2.6 \pm 0.5$ & $16 \pm 2$ & 0.16 & 1.276\\
 $53$\cite{Alper1} & & $15 \pm 2$ & $2.4 \pm 0.2$ & $15 \pm 1$ & 0.16 & 1.722\\
 $63$\cite{Alper1} & & $15 \pm 1$ & $1.6 \pm 0.2$ & $11 \pm 1$ & 0.15 & 1.831\\
\hline
\end{tabular}
\end{ruledtabular}
\end{table}

\begin{table}
\begin{ruledtabular}
\caption{Fit Values of $p_0$ and $n$ for $P\bar{P} \rightarrow (K^++K^-)/2 + X$ at different energies}
\begin{tabular}{ccccccc}
\hline $\sqrt{s}_{NN}(GeV)$ & Relevant & $C_1$ & $p_0$(GeV/c) & $n$ &
$p_0/n$(GeV/c) & $\chi^2/ndf$\\
$[$with reference $]$ & collision-specifics & & & & & \\
\hline $300$\cite{Alexopoulos1} & & $5.7\pm 0.5$ & $1.5\pm 0.3$ & $8.2\pm 0.5$ & 0.18 & 2.258\\
 $540$\cite{Alexopoulos1} & $|y_{cm}|<2.0$, & $4.1 \pm 0.1$ & $1.3\pm0.4$ & $8 \pm 2$ & 0.16 & 2.273\\
 $1000$\cite{Alexopoulos1} & min. bias & $0.36 \pm 0.02$ & $1.50\pm 0.06$ & $7.0\pm 0.2$ & 0.21 & 1.971\\
 $1800$\cite{Alexopoulos1} & & $0.09 \pm 0.01$ & $1.55\pm 0.05$ & $ 8\pm 2$ & 0.19 & 3.526\\
\hline
\end{tabular}
\end{ruledtabular}
\end{table}

\begin{table}
\begin{ruledtabular}
\caption{Fit Values of $p_0$ and $n$ for $PP \rightarrow P + X$ at different energies}
\begin{tabular}{ccccccc}
\hline $\sqrt{s}_{NN}(GeV)$ & Relevant & $C_1$ & $p_{0}$(GeV/c) & $n$ & $p_0/n$(GeV/c) & $\chi^2/ndf$\\
$[$with reference$]$ & collision-specifics & & & & & \\
\hline $23$\cite{Alper1} & & $11 \pm 2$ & $19 \pm 6$ & $86 \pm 15$ & 0.22 & 0.984 \\
 $31$\cite{Alper1} & & $10 \pm 2$ & $16 \pm 5$ & $70 \pm 10$ & 0.23 & 1.012\\
 $45$\cite{Alper1} & $y_{cm}=0$,& $9 \pm 1$ & $12 \pm 3$ & $50 \pm 9$ & 0.24 & 1.359 \\
 $53$\cite{Alper1} & min. bias & $10 \pm 3$ & $8 \pm 1$ & $37 \pm 6$ & 0.22 & 1.022\\
 $63$\cite{Alper1} & & $10 \pm 1$ & $7 \pm 3$ & $34 \pm 7$ & 0.21 & 2.106\\
\hline
\end{tabular}
\end{ruledtabular}
\end{table}

\begin{table}
\begin{ruledtabular}
\caption{Fit Values of $p_0$ and $n$ for $PP \rightarrow \bar{P} + X$ at different energies}
\begin{tabular}{ccccccc}
\hline $\sqrt{s}_{NN}(GeV)$ & Relevant & $C_1$ & $p_{0}$(GeV/c) & $n$ & $p_0/n$(GeV/c) & $\chi^2/ndf$\\
$[$with reference $]$  & collision-specifics & & & & & \\
\hline $23$\cite{Alper1} & & $4.2 \pm .4$ & $15 \pm 3$ & $69 \pm 9$ & 0.22 & 1.085 \\
 $31$\cite{Alper1} & & $5.2 \pm .5$ & $13 \pm 2$ & $60 \pm 11$ & 0.22 & 2.104\\
 $45$\cite{Alper1} & $y_{cm}=0$, & $7 \pm 0.4$ & $11 \pm 4$ & $53 \pm 8$ & 0.21 & 1.734 \\
 $53$\cite{Alper1} & min. bias & $6.6 \pm 0.7$ & $10 \pm 2$ & $45 \pm 9$ & 0.22 & 1.430\\
 $63$\cite{Alper1} & & $9.3 \pm 0.4$ & $9 \pm 3$ & $40 \pm 5$ & 0.23 & 1.226\\
\hline
\end{tabular}
\end{ruledtabular}
\end{table}

\begin{table}
\begin{ruledtabular}
\caption{Fit Values of $p_0$ and $n$ for $P\bar{P} \rightarrow \bar{P} + X$ at different energies}
\begin{tabular}{ccccccc}
\hline $\sqrt{s}_{NN}(GeV)$ & Relevant & $C_1$ & $p_0$(GeV/c) & $n$ &
$p_0/n$(GeV/c) & $\chi^2/ndf$\\
$[$with reference$]$ & collision-specifics & & & & & \\
\hline $300$\cite{Alexopoulos1} & & $2.6\pm 0.2$ & $7.5\pm 0.9$ & $29\pm 3$ & 0.26 & 1.242\\
 $540$\cite{Alexopoulos1} & $|y_{cm}|< 2.0$, & $1.4 \pm 0.1$ & $7\pm 2$ & $25 \pm 6$ & 0.28 & 1.775\\
 $1000$\cite{Alexopoulos1} & min. bias & $0.13 \pm 0.03$ & $7.3\pm 0.8$ & $24 \pm 3$ & 0.30 & 1.620\\
 $1800$\cite{Alexopoulos1} & & $0.025 \pm 0.003$ & $7.4\pm 0.9$ & $23 \pm 2$ & 0.32 & 1.483\\
\hline
\end{tabular}
\end{ruledtabular}
\end{table}

\begin{table}
\begin{ruledtabular}
\caption{Values of $a,b,\acute{a}, \acute{b}$}
\begin{tabular}{ccccc}
\hline Secondaries & $a$ & $b$ & $\acute{a}$ &
$\acute{b}$ \\
\hline $(K^++K^-)/2$ & $1.6$ & $103$ & $3.6$ & 161\\
 $P$ & $7$ & $602$ & $5$ & 644\\
 $\bar{P}$ & $7$ & $478$ & $13$ & 527\\
\hline
\end{tabular}
\end{ruledtabular}
\end{table}

\begin{table}
\begin{ruledtabular}
\caption{Numerical Values of the parameters: $\alpha$ and $\beta$ for Kaon
production in different high energy collisions}
\begin{tabular}{ccccccc}
\hline  ${\rm Collision}$ &  $E$(GeV) & Relevant & $C_3$ &  $\alpha$ & $\beta$ & $\chi^2/ndf$\\
$[$with reference$]$ & & collision-specifics & & (GeV/c)$^{-1}$ & (GeV/c)$^{-2}$ & \\
\hline ${\rm P+D}$\cite{Antreasyan1} &  $400$ & $y=0$, min. bias &  $38 \pm 6$ & $0.14 \pm 0.03 $ & $0.04 \pm 0.01$ & 2.832\\
 ${\rm P+Be}$\cite{Boggild1} &  $450$ & $2.4<y<3.5$, central & $1.2 \pm 0.4$ & $0.26 \pm 0.03 $ & $0.04 \pm 0.015$ & 0.381 \\
 ${\rm P+S}$\cite{Baechler1} &  $200$ & $0.6<y<2.4$, central & $6.4 \pm 0.6$ & $0.23 \pm 0.03$ &  $0.03 \pm 0.01$ & 1.974\\
 ${\rm P+Au}$\cite{Baechler1} &  $200$ & $0.2<y<2.0$, central & $6.3 \pm 0.4$ & $0.21 \pm 0.03$ &  $0.032 \pm 0.005$ & 0.859\\
 ${\rm P+Pb}$\cite{Boggild1} &  $450$ & $2.4<y<3.5$, central & $1.1 \pm 0.1$ & $0.17 \pm 0.04$ & $0.04 \pm 0.01$ & 0.552\\
 ${\rm O+Au}$\cite{Baechler1} & $200A$ & $0.2<y<2.0$, central & $63 \pm 9$ & $0.24 \pm 0.05$ &  $0.03 \pm 0.01$ & 1.164\\
 ${\rm S+S}$\cite{Baechler1} & $200A$ & $0.6<y<2.4$, central & $72 \pm 10$ & $0.24 \pm 0.02$ &  $0.03 \pm 0.01$ & 2.088\\
 ${\rm S+Pb}$\cite{Boggild1} & $200A$ & $2.3<y<2.9$, central & $28\pm 3$ &  $0.25 \pm 0.03$ &  $0.027 \pm 0.002$ & 2.170\\
 ${\rm Au+Au}$\cite{Velkovska1} & $8450A$ & $|\eta_{cm}|<0.35$, min. bias & $17\pm 2$ & $0.15 \pm 0.03$ &  $0.041 \pm 0.005$ & 1.005 \\
 ${\rm Pb+Pb}$\cite{Bearden2} &  $160A$ & $2.4<y<3.5$, central & $66 \pm 2$ & $0.20 \pm 0.02$ & $0.038 \pm 0.004$ & 2.158\\
\hline
\end{tabular}
\end{ruledtabular}
\end{table}

\begin{table}
\begin{ruledtabular}
\caption{Numerical Values of the parameters: $\alpha$ and $\beta$ for
Proton production in different high energy collisions}
\begin{tabular}{ccccccc}
\hline  ${\rm Collision}$ &  $E$(GeV) & Relevant & $C_3$ &  $\alpha$ & $\beta$ & $\chi^2/ndf$\\
$[$with reference$]$ & & collision-specifics & & (GeV/c)$^{-1}$ & (GeV/c)$^{-2}$ & \\
 \hline ${\rm P+D}$\cite{Antreasyan1} &  $400$ & $y=0$, min. bias & $28 \pm 2$ & $0.18 \pm 0.02 $ & $0.028 \pm 0.002$ & 2.577\\
 ${\rm P+Be}$\cite{Bearden1} &  $450$ & $2.3<y<2.9$, central & $0.57 \pm 0.02$ & $0.22 \pm 0.04$ & $0.04 \pm 0.01$ & 2.530\\
 ${\rm P+S}$\cite{Alber1} &  $200$ & $0.5<y<3.0$, min. bias & $16 \pm 2$ & $0.15 \pm 0.02$ &  $0.045 \pm 0.008$ & 1.813\\
 ${\rm P+Au}$\cite{Alber1} &  $200$ & $0.5<y<3.0$, min. bias & $47 \pm 6$ & $0.14 \pm 0.03$ &  $0.05 \pm 0.02$ & 1.425\\
 ${\rm P+Pb}$\cite{Bearden1} &  $450$ & $2.3<y<2.9$, central & $0.8\pm 0.2$ & $0.18 \pm 0.02$ & $0.04 \pm 0.01$ & 2.763\\
 ${\rm D+Au}$\cite{Alber1} &  $200A$ & $0.5<y<3.0$, min. bias &  $102 \pm 8$ & $0.16 \pm 0.02$ &  $0.05 \pm 0.02$ & 0.535\\
 ${\rm O+Au}$\cite{Alber1} & $200A$ & $0.5<y<3.0$, min. bias & $473 \pm 63$ & $0.17 \pm 0.01$ &  $0.04 \pm 0.01$ & 0.866\\
 ${\rm S+S}$\cite{Alber1} & $200A$ & $0.5<y<3.0$, min. bias & $164 \pm 30$ & $0.19 \pm 0.02$ &  $0.03 \pm 0.01$ & 1.247\\
 ${\rm S+Ag}$\cite{Alber1} & $200A$ & $0.5<y<3.0$, min. bias & $450 \pm 70$ & $0.17 \pm 0.01$ &  $0.03 \pm 0.01$ & 0.992\\
 ${\rm S+Au}$\cite{Alber1} &  $200A$ & $3.0<y<5.0$, min. bias & $139\pm 12$ & $0.17\pm 0.03$ &  $0.03\pm 0.01$ & 2.063\\
 ${\rm Au+Au}$\cite{Velkovska1} & $8450A$ & $|\eta_{cm}|<0.35$, min. bias & $7\pm 2$ & $0.17 \pm 0.02$ &  $0.034 \pm 0.001$ & 0.744 \\
 ${\rm Pb+Pb}$\cite{Bearden2} &  $160A$ & $2.3<y<2.9$, central & $33\pm 2$ & $0.19 \pm 0.02$ & $0.022 \pm 0.002$ & 2.672\\
\hline
\end{tabular}
\end{ruledtabular}
\end{table}

\begin{table}
\begin{ruledtabular}
\caption{Numerical Values of the parameters: $\alpha$ and $\beta$ for
Antiproton production in different high energy collisions}
\begin{tabular}{ccccccc}
\hline  ${\rm Collision}$ &  $E$(GeV) & Relevant & $C_3$ &  $\alpha$ & $\beta$ & $\chi^2/ndf$\\
$[$with reference$]$ & & collision-specifics & & (GeV/c)$^{-1}$ & (GeV/c)$^{-2}$ &\\
\hline ${\rm P+D}$\cite{Antreasyan1} & $400$ & $y_{cm}=0$, min. bias &  $7.2 \pm 0.6$ & $0.15 \pm 0.04 $ & $0.035 \pm 0.003$& 1.764\\
 ${\rm P+Be}$\cite{Bearden1} &  $450$ & $2.3<y<2.9$, central & $0.16 \pm 0.03$ & $0.20 \pm 0.03$ & $0.03 \pm 0.01$ & 1.362\\
 ${\rm P+Pb}$\cite{Bearden1} &  $450$ & $2.3<y<2.9$, central & $0.18\pm .04$ & $0.19 \pm 0.03$ & $0.032 \pm 0.004$ & 1.470\\
 ${\rm S+S}$\cite{Bearden1} & $200A$ & $2.3<y<2.9$, central & $1.5 \pm 0.2$ & $0.19 \pm 0.02$ &  $0.03 \pm 0.01$ & 2.674\\
 ${\rm S+Ag}$\cite{Alber2} & $200A$ & $3<y<4$, central & $8 \pm 1$ & $0.18 \pm 0.02$ &  $0.03 \pm 0.01$ & 1.851\\
 ${\rm S+Au}$\cite{Alber2} &  $200A$ & $3<y<4$, central & $5.7\pm 0.6$ & $0.25\pm 0.03$ &  $0.024\pm 0.003$ & 1.620\\
 ${\rm S+Pb}$\cite{Bearden1} & $200A$ & $2.3<y<2.9$, central & $1.7 \pm 0.2$ &  $0.23 \pm 0.03$ &  $0.022 \pm 0.003$ & 1.767\\
 ${\rm Au+Au}$\cite{Velkovska1} & $8450A$ & $|\eta_{cm}|<0.35$, min. bias & $0.42\pm 0.01$ & $0.16 \pm 0.03$ &  $0.034 \pm 0.003$ & 2.081\\
 ${\rm Pb+Pb}$\cite{Bearden2} &  $160A$ & $2.3<y<2.9$, central &  $15 \pm 1$ & $0.26 \pm 0.02$ & $0.024 \pm 0.002$ & 2.838\\
\hline
\end{tabular}
\end{ruledtabular}
\end{table}

\newpage

\begin{figure}
\subfigure[]{
\begin{minipage}{.5\textwidth}
\centering
\includegraphics[width=7cm]{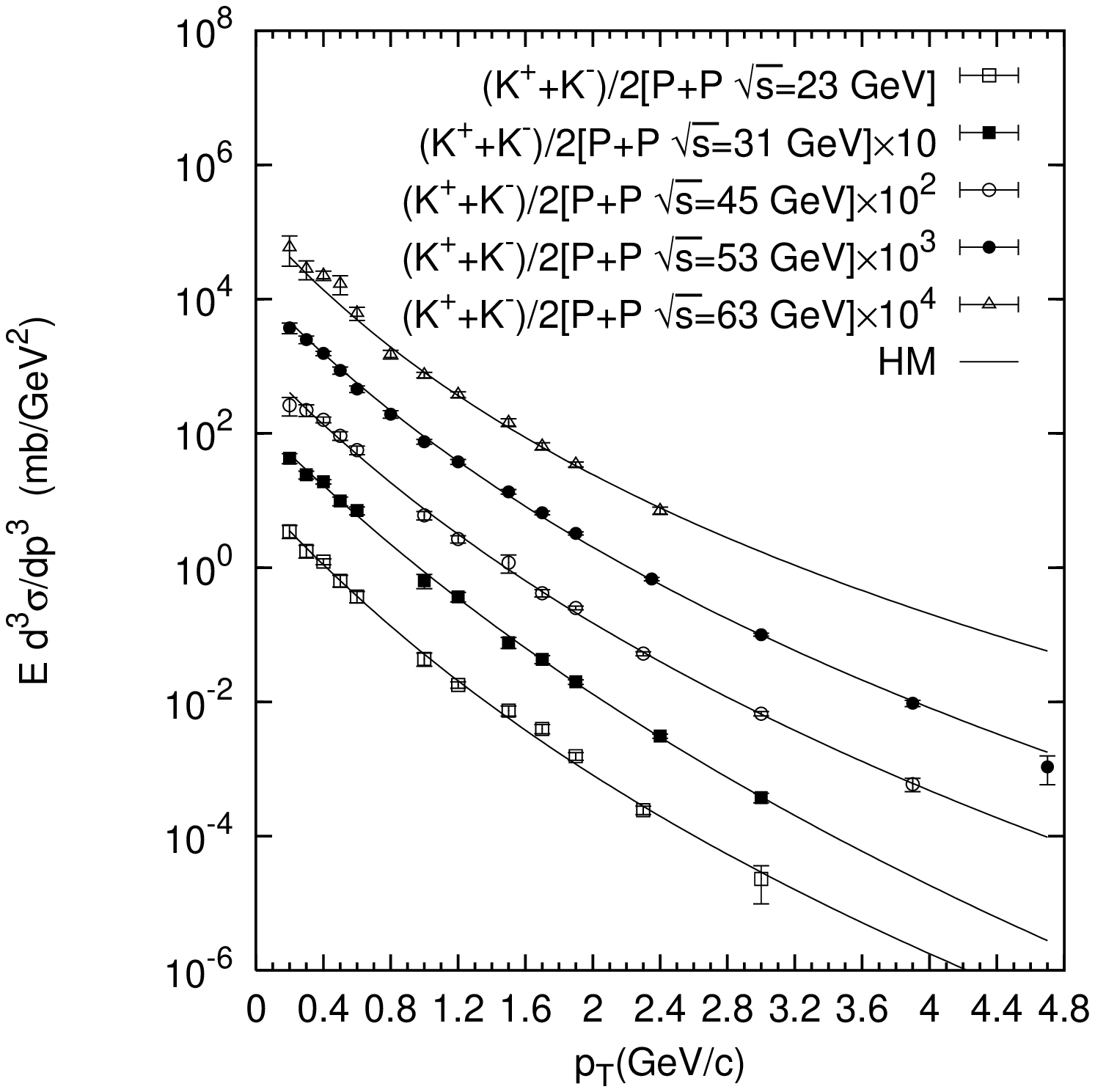}
\end{minipage}}%
\subfigure[]{
\begin{minipage}{.5\textwidth}
\centering
\includegraphics[width=7cm]{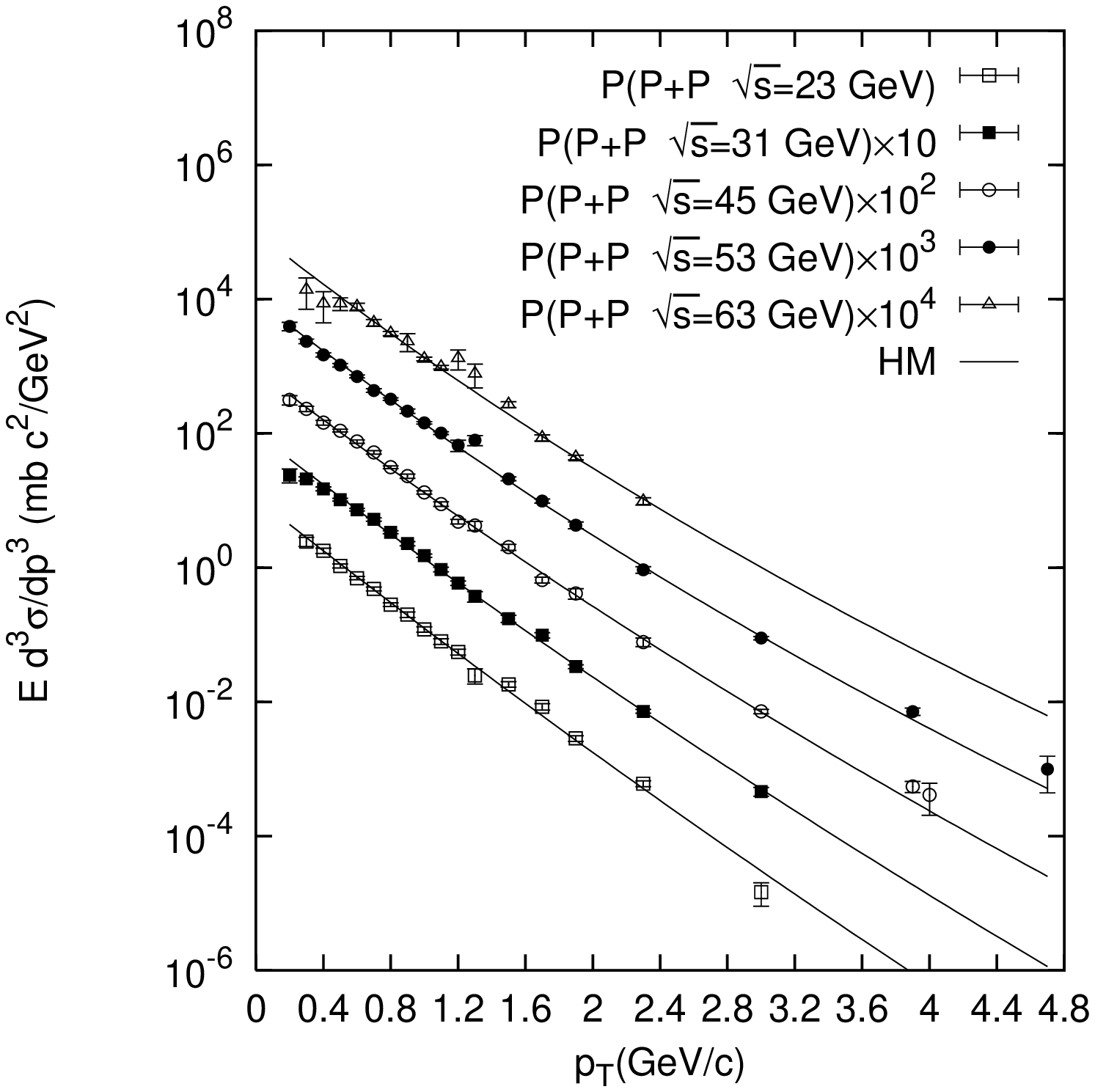}
\end{minipage}}%
\vspace{.1cm}
\subfigure[]{
\begin{minipage}{1\textwidth}
\centering
\includegraphics[width=7cm]{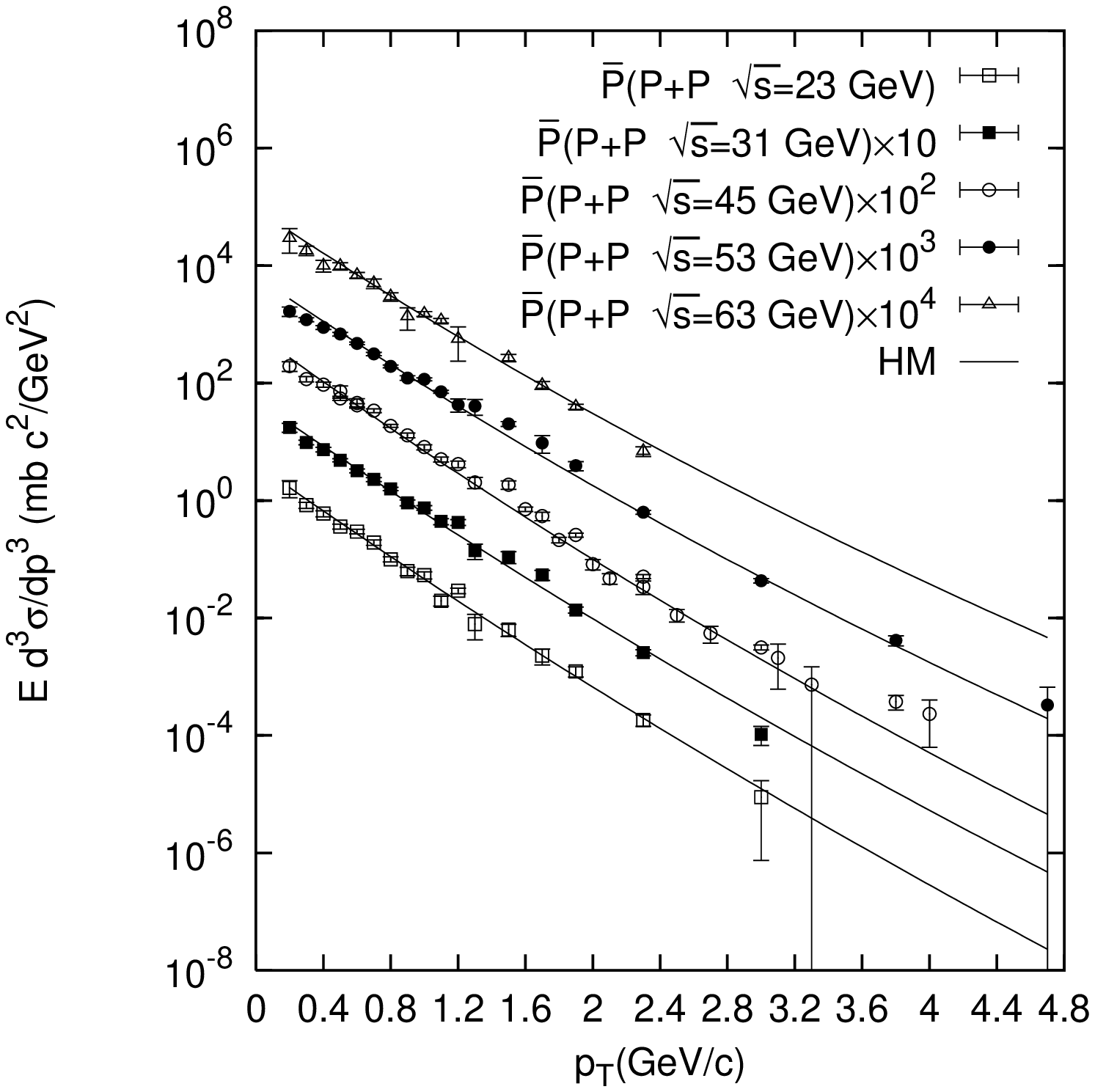}
\end{minipage}}%
\caption{Plot of $E\frac{d^3\sigma}{dp^3}$ vs. $p_T$ for various
non-pionic secondaries produced in $P+P$ collisions at different
c.m. energies. The various experimental points are taken from
Ref.\cite{Alper1}. The solid curves give the theoretical fits on
the basis of Hagedorn's model(eqn.3).}
\end{figure}
\begin{figure}
\subfigure[]{
\begin{minipage}{.5\textwidth}
\centering
\includegraphics[width=7cm]{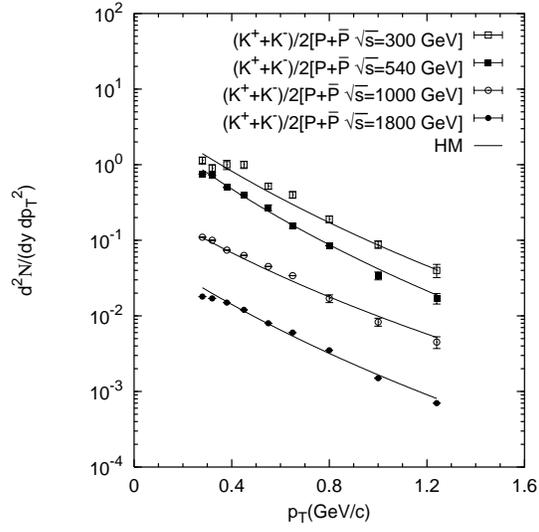}
\end{minipage}}%
\subfigure[]{
\begin{minipage}{.5\textwidth}
\centering
\includegraphics[width=7cm]{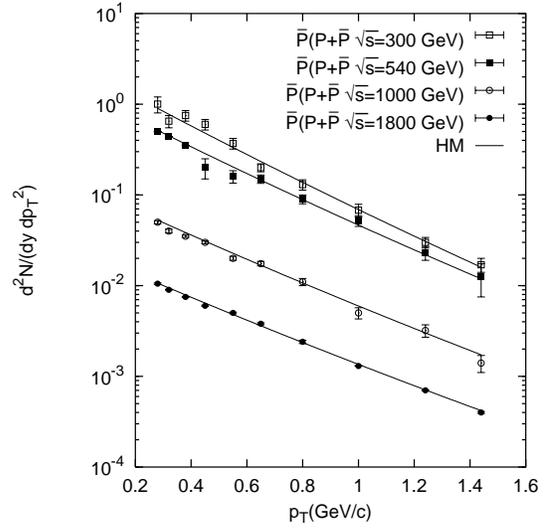}
\end{minipage}}%
\caption{The inclusive spectra for secondary $\frac{K^++K^-}{2}$
and $\bar{P}$ produced in $P+\bar{P}$ collisions at
$\sqrt{s}=300,540,1000$ and $1800$ GeV. The various experimental
points are from Ref.\cite{Alexopoulos1}. The solid curvilinear
lines are drawn on the basis of eqn.3.}
\end{figure}
\begin{figure}
\subfigure[]{
\begin{minipage}{.5\textwidth}
\centering
\includegraphics[width=7cm]{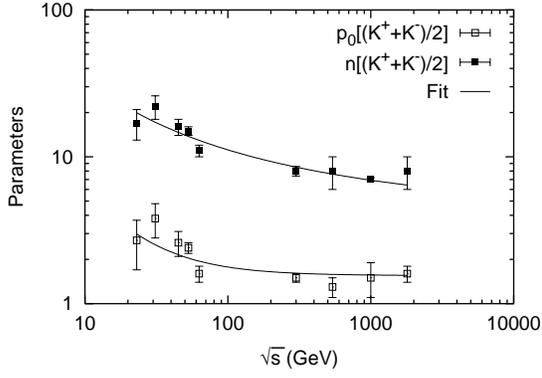}
\end{minipage}}%
\subfigure[]{
\begin{minipage}{.5\textwidth}
\centering
\includegraphics[width=7cm]{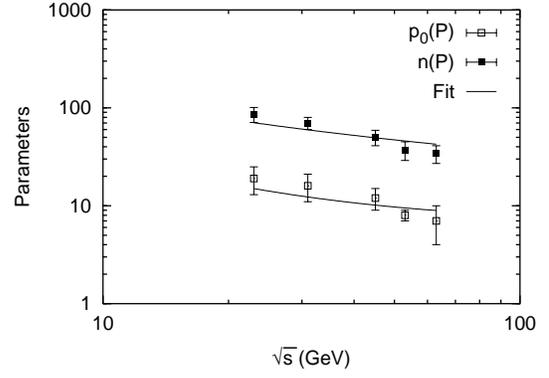}
\end{minipage}}%
\vspace{.1cm}
\subfigure[]{
\begin{minipage}{1\textwidth}
\centering
\includegraphics[width=7cm]{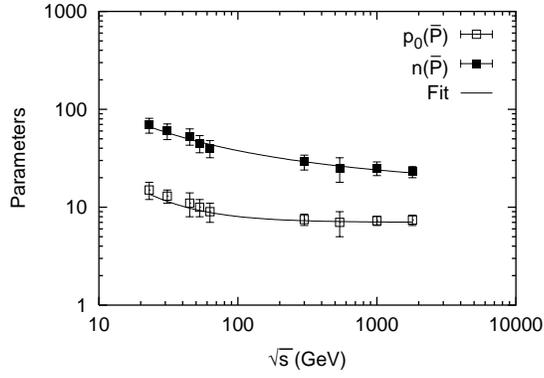}
\end{minipage}}%
\caption{Values of $p_0$ and $n$ as a function of c.m. energy
$\sqrt{s}$. Various data points are taken from Table-I to Table-V.
The solid curves are drawn on the basis of eqn.4 and eqn.5.}
\end{figure}
\begin{figure}
\begin{minipage}{.5\textwidth}
\centering
\includegraphics[width=7cm]{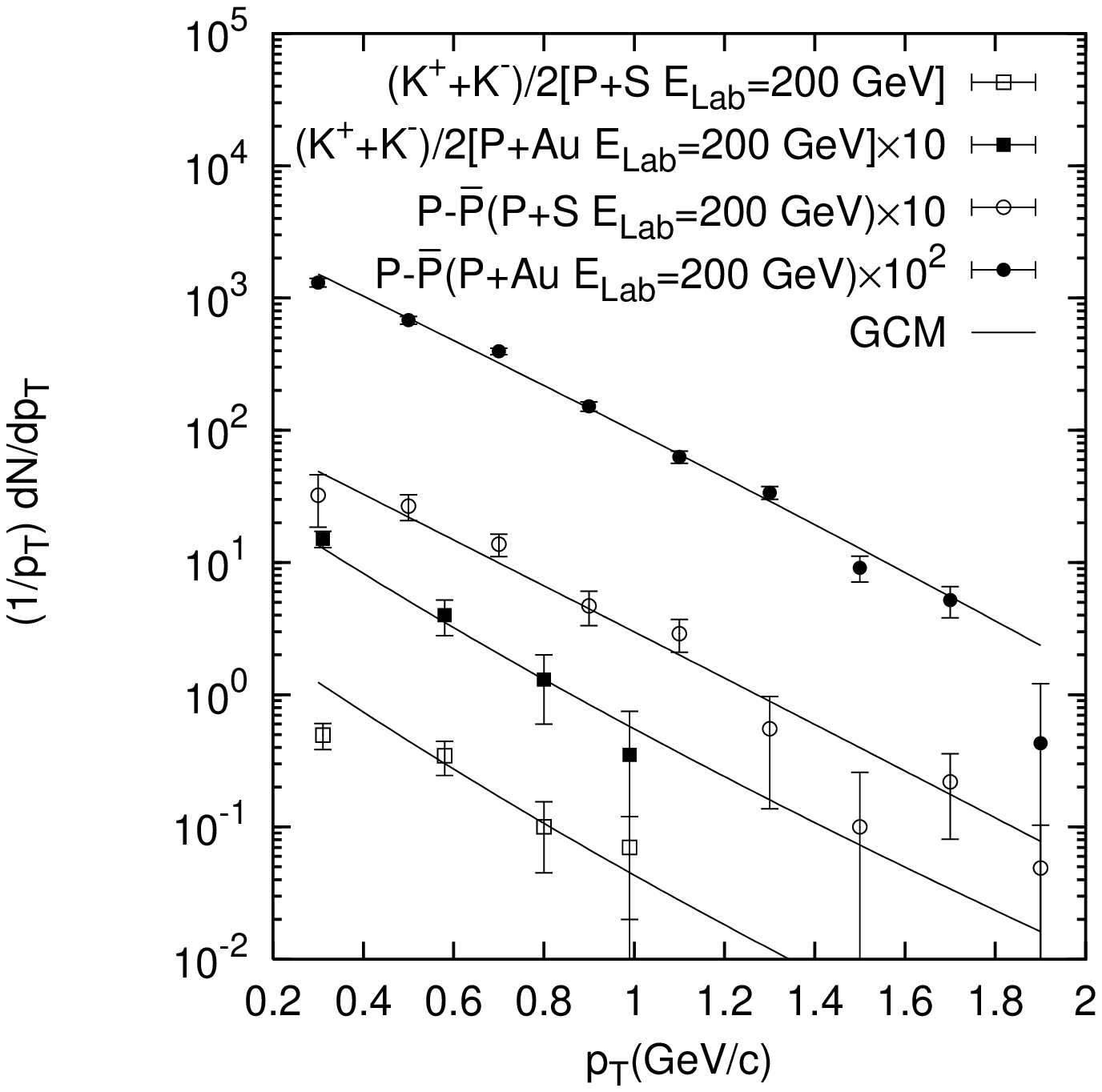}
\setcaptionwidth{7cm} \caption{Plot of
$(\frac{1}{p_T})\frac{dN}{dp_T}$ as a function of transverse
momentum, $p_T$ for production of kaons and net protons in $PS$
and $PAu$ collisions at $E_{Lab}=200$ GeV. The various
experimental points are taken from Ref.\cite{Baechler1,Alber1}.
The solid curves depict the GCM-based results(eqn.8).}
\end{minipage}%
\begin{minipage}{.5\textwidth}
\centering
\includegraphics[width=7cm]{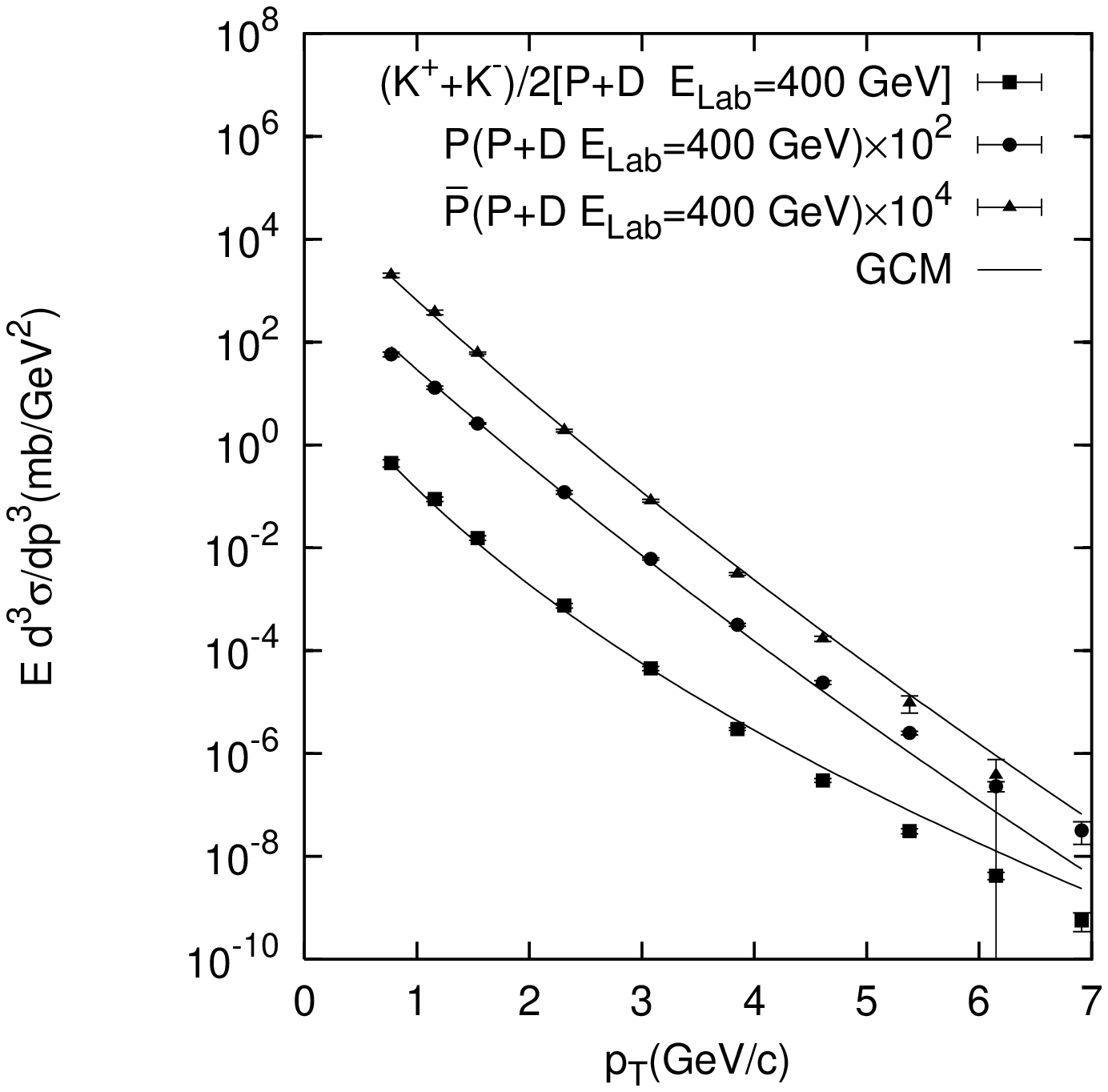}
\setcaptionwidth{7cm} \caption{Plot of $E\frac{d^3\sigma}{dp^3}$
vs. $p_T$ for various secondaries produced in $P+D$ collisions at
$E_{Lab}=400$ GeV. The various experimental points are taken from
Ref.\cite{Antreasyan1}. The solid curves provide the fits on the
basis of eqn.8.}
\end{minipage}%
\end{figure}
\begin{figure}
\centering
\includegraphics[width=7cm]{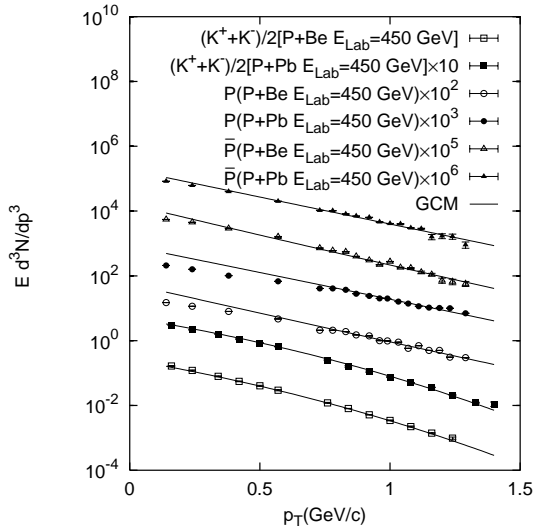}
\caption{Nature of Invariant spectra for production of
$\frac{K^++K^-}{2}$, $P$ and $\bar{P}$ in $P+Be$ and $P+Pb$
collisions at $E_{Lab}=450$ GeV. Various measured data are
obtained from Ref.\cite{Boggild1,Bearden1}. The solid curvilinear
lines are drawn on the basis of GCM.}
\end{figure}
\begin{figure}
\subfigure[]{
\begin{minipage}{.5\textwidth}
\centering
\includegraphics[width=7cm]{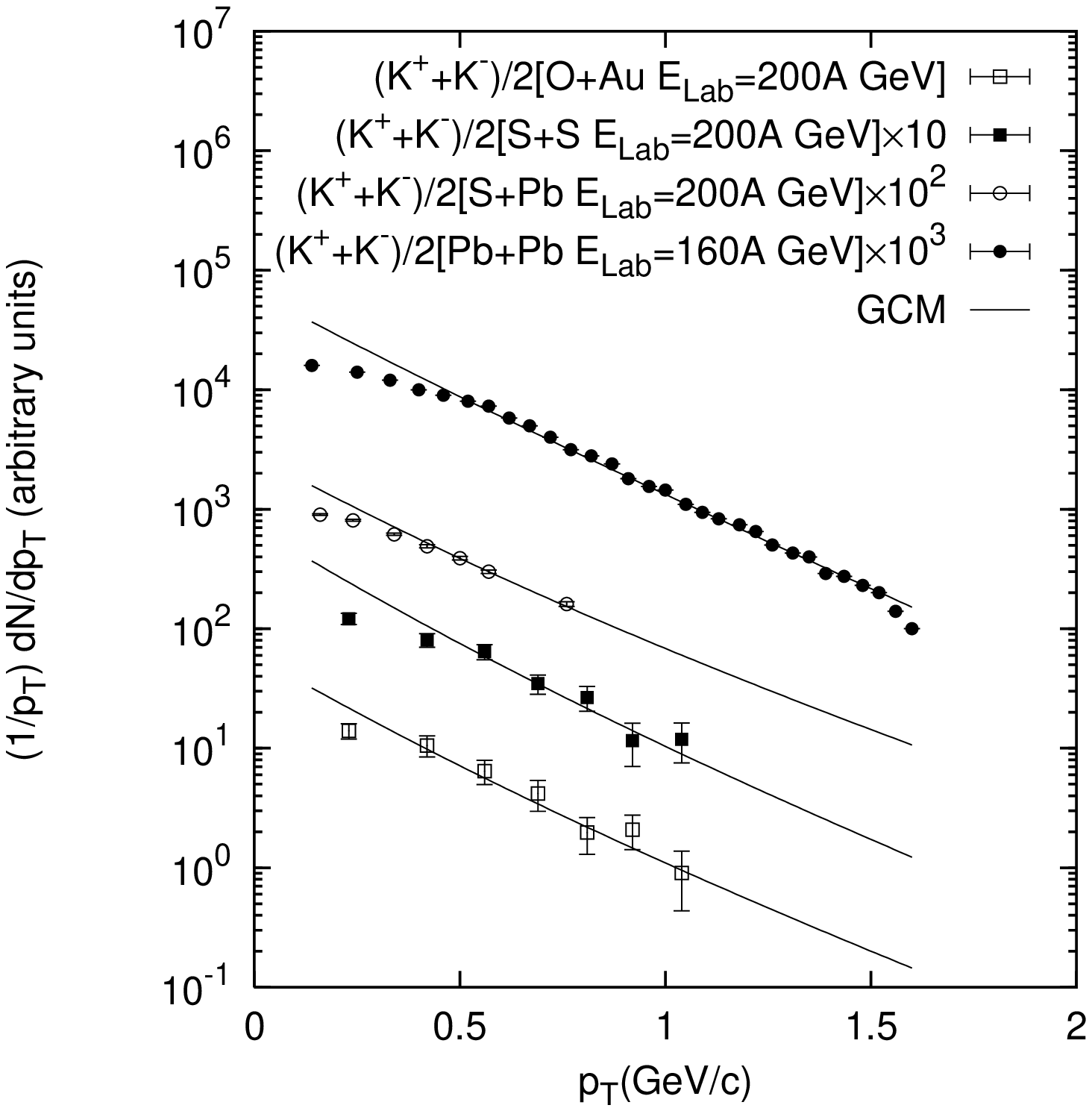}
\end{minipage}}%
\subfigure[]{
\begin{minipage}{.5\textwidth}
\centering
\includegraphics[width=7cm]{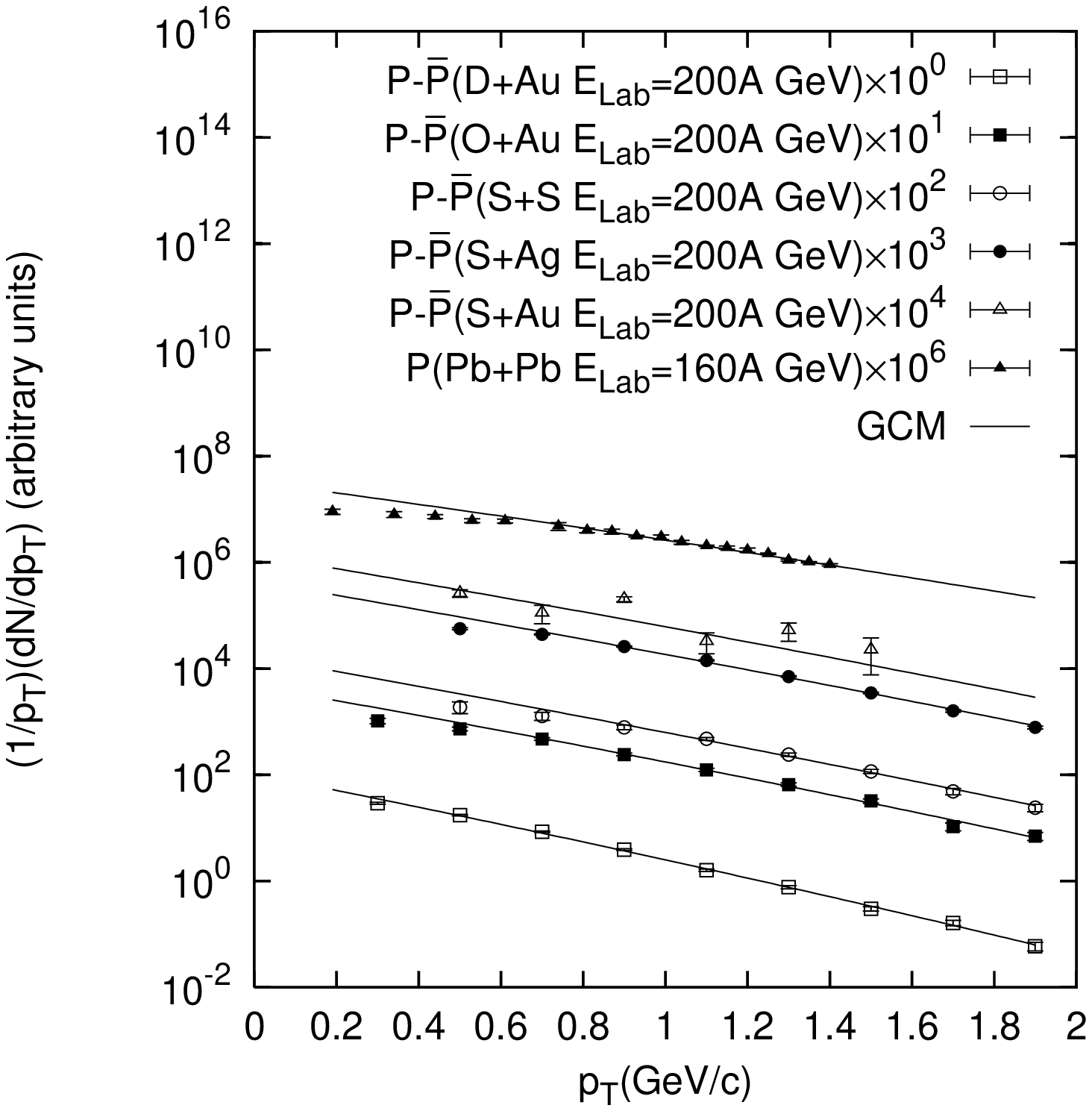}
\end{minipage}}%
\vspace{.1cm}
\subfigure[]{
\begin{minipage}{1\textwidth}
\centering
\includegraphics[width=7cm]{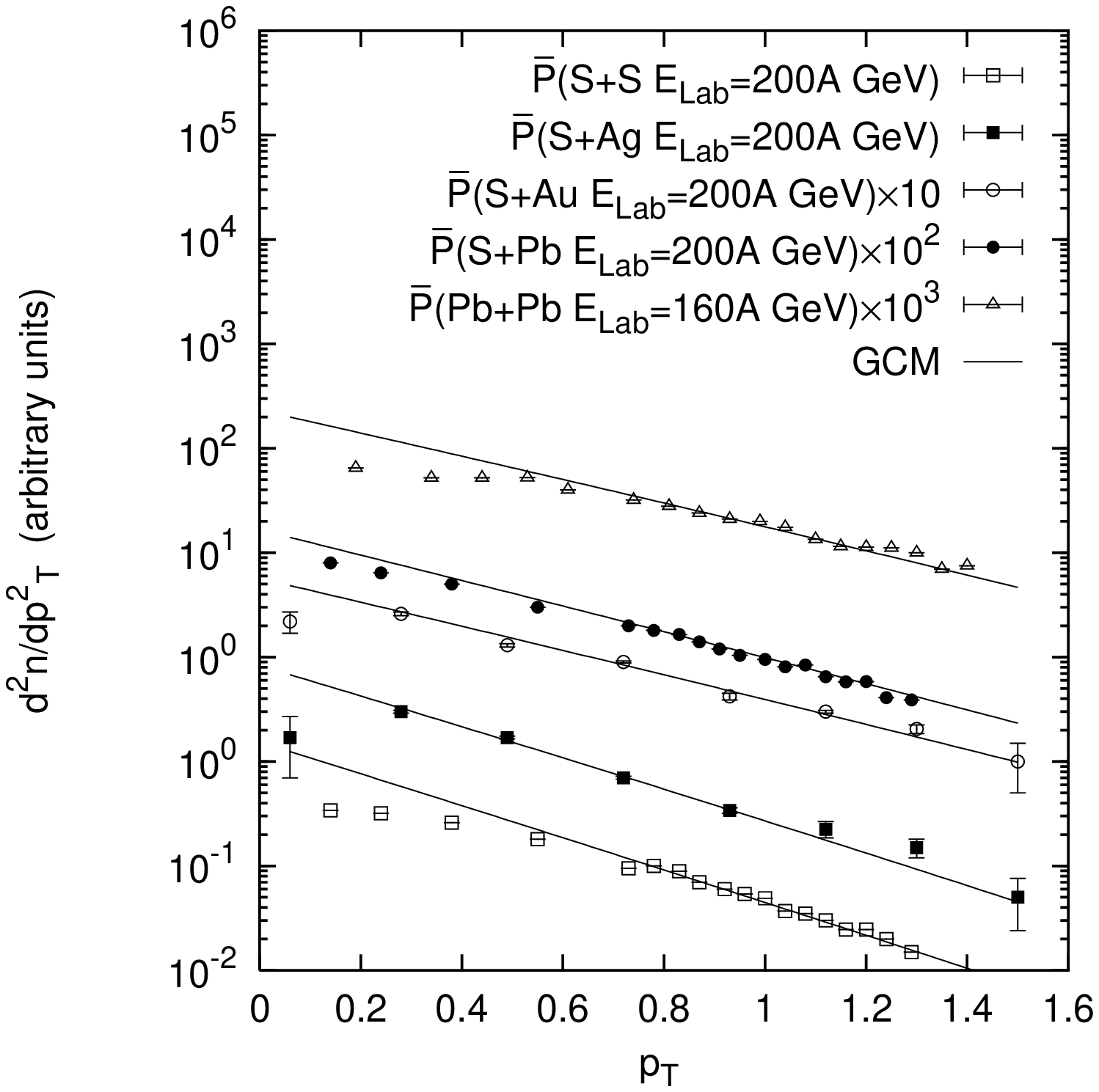}
\end{minipage}}%
\caption{Plots of multiplicity spectra as function of $p_T$ for
production of some non-pionic secondaries in different
nucleus-nucleus collisions at CERN and SPS energies. Various
experimental data are taken from
Ref.\cite{Baechler1,Alber1,Boggild1,Bearden1,Bearden2,Alber2}. The
solid curves depict the GCM-based fits(eqn.8).}
\end{figure}
\begin{figure}
\centering
\includegraphics[width=7cm]{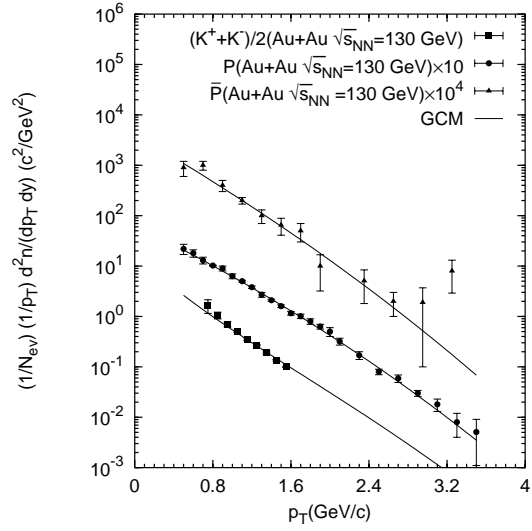}
\caption{The nature of invariant spectra for secondary charged
kaons, protons and antiprotons produced in $AuAu$ interactions at
RHIC energy($\sqrt{s_{NN}}=130$ GeV). The experimental data are
from Ref.\cite{Velkovska1}. The solid curves provide the present
model-based fits. }
\end{figure}
\begin{figure}
\subfigure[]{
\begin{minipage}{.5\textwidth}
\centering
\includegraphics[width=7cm]{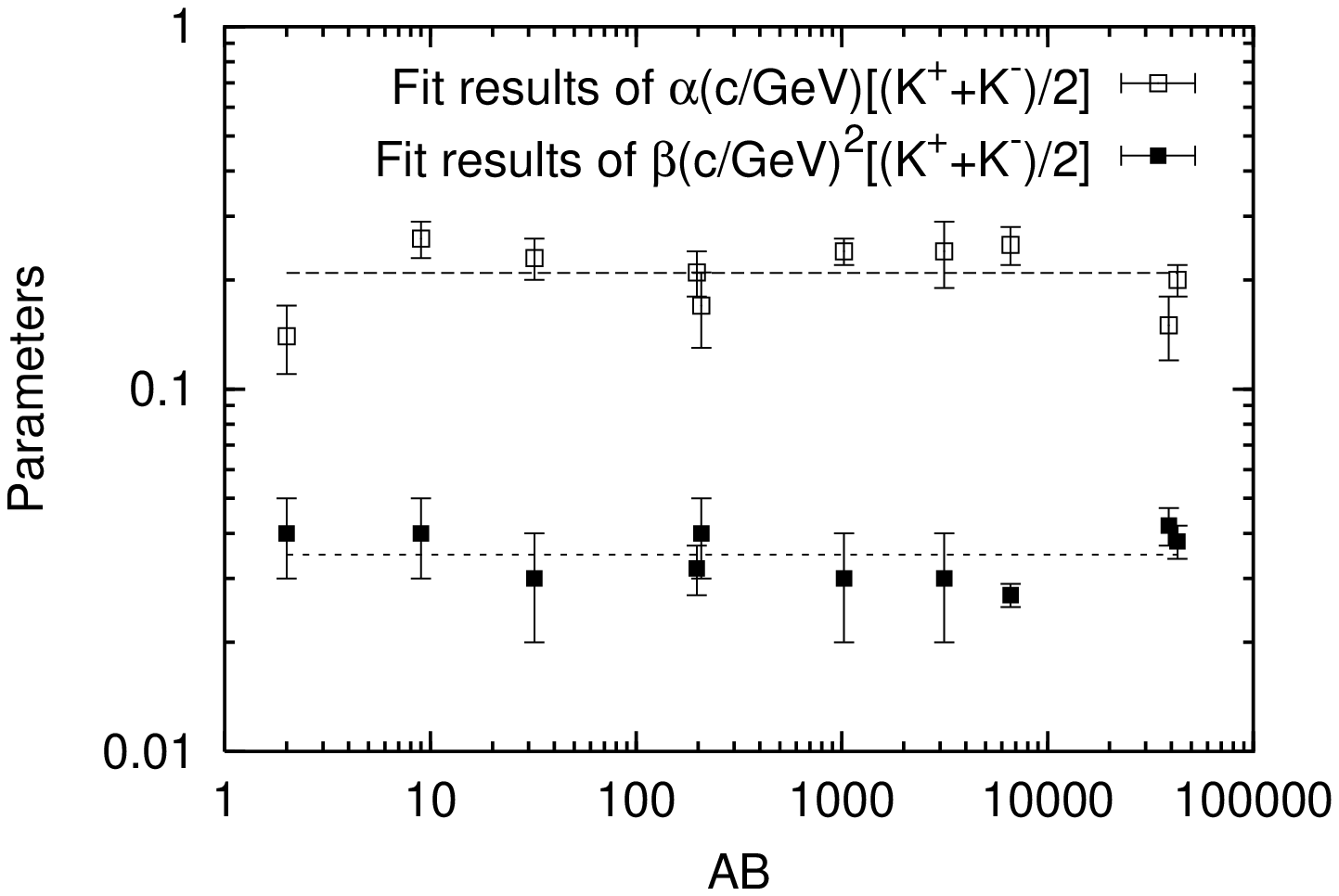}
\end{minipage}}%
\subfigure[]{
\begin{minipage}{.5\textwidth}
\centering
\includegraphics[width=7cm]{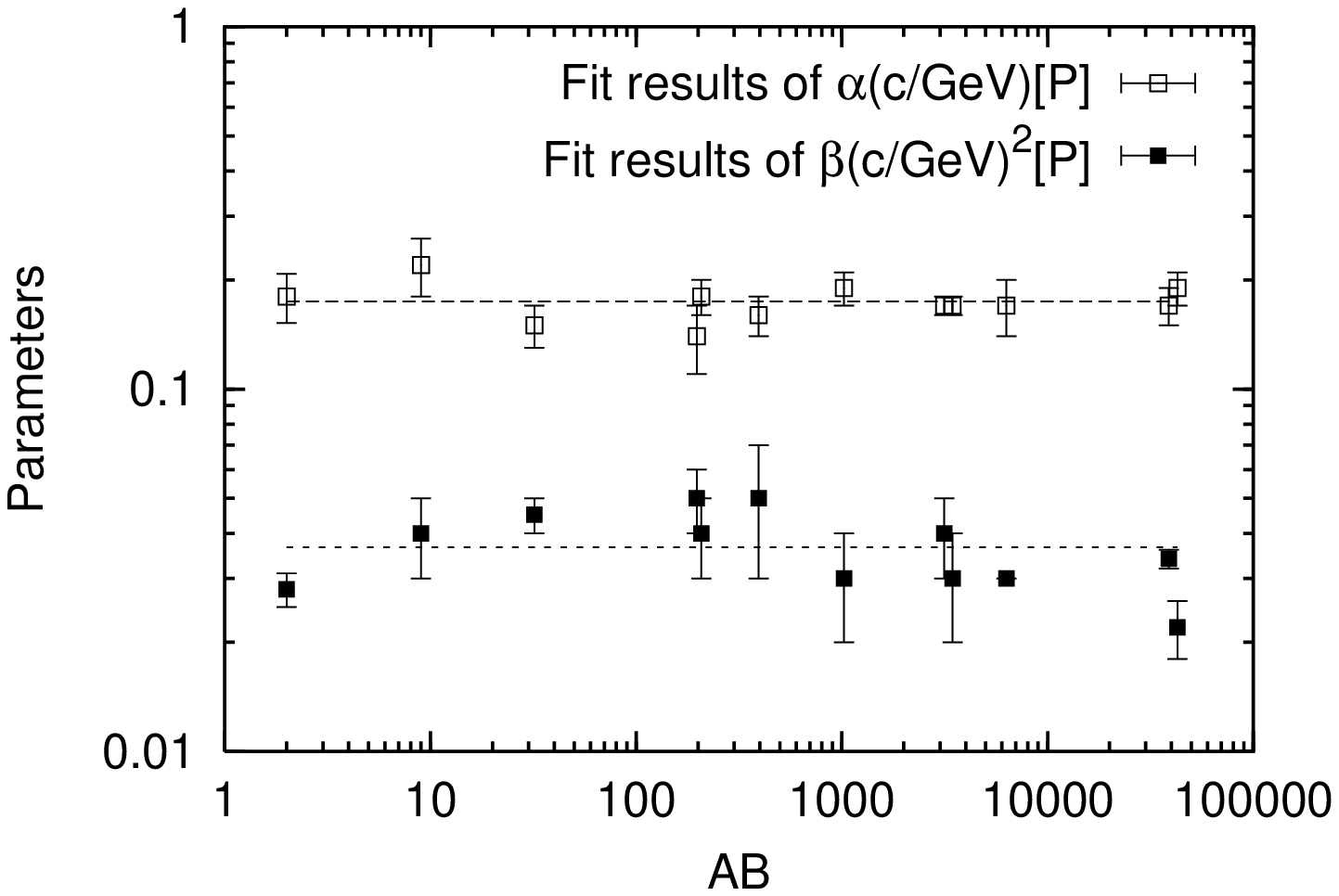}
\end{minipage}}%
\vspace{.1cm}
\subfigure[]{
\begin{minipage}{1\textwidth}
\centering
\includegraphics[width=7cm]{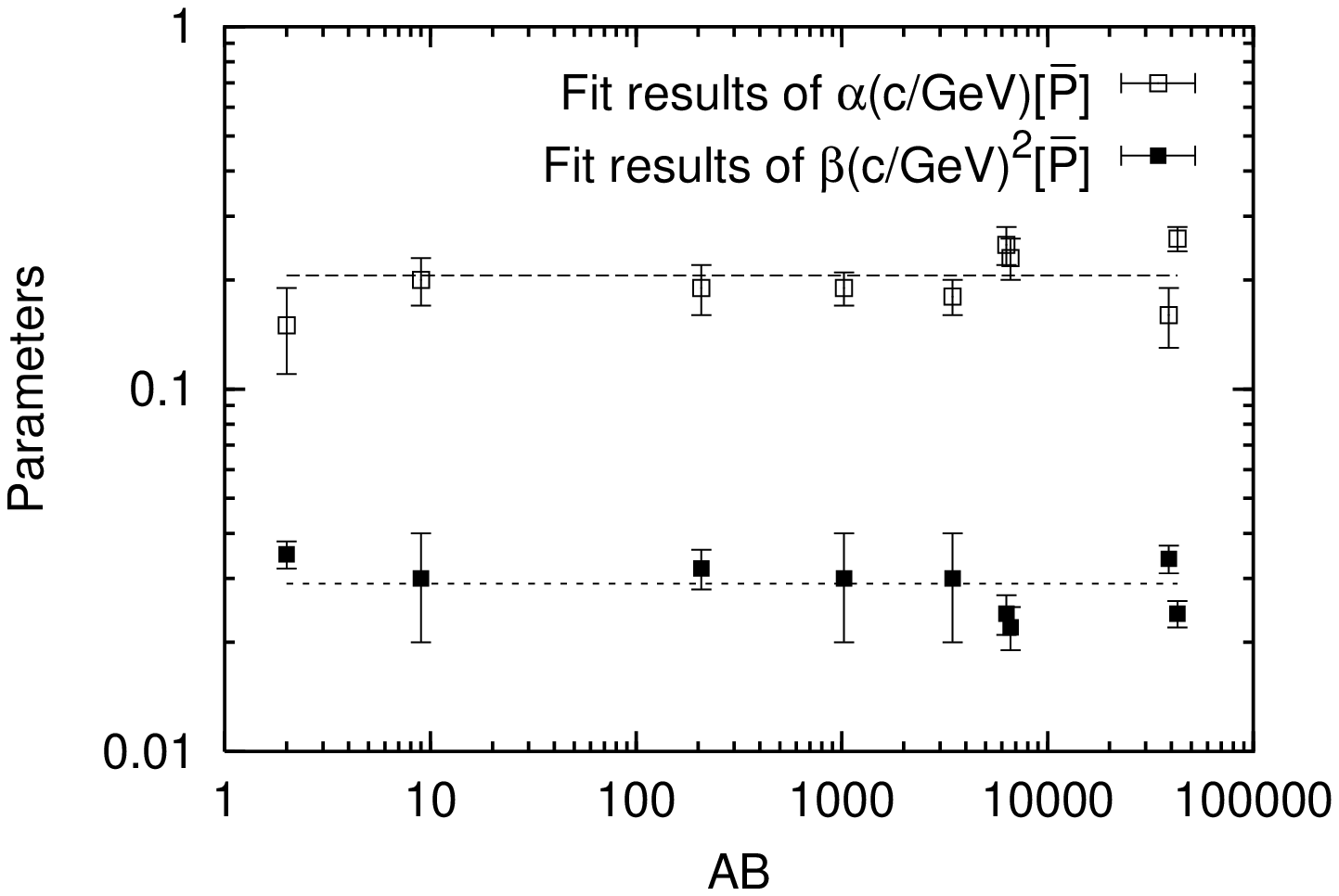}
\end{minipage}}%
\caption{Values of $\alpha$ and $\beta$ for various non-pionic
secondaries produced in different collisions as functions of the
product of mass numbers($AB$) of the interacting nuclei. The
fitted values of $\alpha$ and $\beta$, enlisted in Table-VII to
Table-IX, are taken as the data points; and are denoted by empty
and filled squares respectively. The dashed lines give the average
values for $\alpha_{\frac{K^++K^-}{2}}=0.21\pm0.03$,
$\alpha_{P}=0.17\pm0.03$, $\alpha_{\bar{P}}=0.21\pm0.04$,
$\beta_{\frac{K^++K^-}{2}}=0.035\pm0.0005$,
$\beta_{P}=0.037\pm0.003$ and $\beta_{\bar{P}}=0.030\pm0.005$.}
\end{figure}
\end{document}